\documentstyle[epsf]{elsart}
\newcommand{\beq}{\begin{equation}}
\newcommand{\eeq}{\end{equation}}
\newcommand{\ds}{\displaystyle}
\newcommand{\beqar}{\begin{eqnarray}}
\newcommand{\eeqar}{\end{eqnarray}}
\begin{document}
\begin{frontmatter}
\title{\bf Supercooling of rapidly expanding quark-gluon plasma }
\author{E.E. Zabrodin$^{a,b}$}, \author{L.V. Bravina$^{a,b,1}$}, 
\author{L.P. Csernai$^{c}$}, \author{H. St{\"o}cker$^{a}$} and 
\author{W. Greiner$^{a}$}
\address{
$^a$ Institute for Theoretical Physics, University of Frankfurt,\\
Robert-Mayer-Str. 8-10, D-60054 Frankfurt, Germany \\ 
$^b$ Institute for Nuclear Physics, Moscow State University,
119899 Moscow, Russia \\
$^c$ Department of Physics, University of Bergen, All{\'e}gaten 55,
5007 Bergen, Norway\\
$^1$ Alexander von Humboldt Fellow
}

\maketitle


\begin{abstract}
We reexamine the scenario of homogeneous nucleation of the 
quark-gluon plasma produced in ultra-relativistic heavy ion 
collisions. A generalization of the standard nucleation theory to 
rapidly expanding system is proposed. The nucleation rate is derived 
via the new scaling parameter $\lambda_Z$. It is shown that the size 
distribution of hadronic clusters plays an important role in
the dynamics of the phase transition. The longitudinally expanding 
system is supercooled to about $3-6\%$, then it is reheated, and
the hadronization is completed within $6-10$ fm/$c$, i.e. $5-10$
times faster than it was estimated earlier, in a strongly
nonequilibrium way.\\
\end{abstract}

\end{frontmatter}

\newpage

Homogeneous nucleation of quark-gluon plasma (QGP) produced in
ultra-relativistic heavy ion collisions has been subject of 
interest during the last few years. Obviously, the plasma must 
hadronize, but the mechanism of hadronization still remains an
open question. In case of a second order phase transition, 
percolation model calculations are useful,
also for strong supercooling of the plasma during a first order
phase transition, provided the expansion is rapid and the amount
of the plasma converted into hadronic matter is not sufficiently 
large to reheat the growing volume. 

Results of simulations \cite{CsKa92b,CKKZ93} show that the
supercooling of the system, if quenched rapidly into the metastable
or even unstable region, may be as large as $20\%$. Then it is
questionable whether the homogeneous nucleation scheme
is appropriate for the description of the hadronization process.
Other processes which may cause a rapid hadronization of QGP have
been studied recently \cite{CsCs94,CsMi95}.

In the present paper we propose a generalization of the standard
nucleation theory to rapidly expanding systems. If the size     
distribution of hadronic bubbles is taken into account, the
total rate of the plasma conversion turns out to be high enough
to prevent the system from strong supercooling. The last
circumstance makes relevant the application of the homogeneous
nucleation scenario. 

The paper is organized as follows: after a sketch of the nucleation
theory the dynamics of the plasma$-$hadrons phase transition in the
Bjorken hydrodynamic model \cite{Bjor83} is discussed. 
Finally, conclusions will be drawn.

The starting point of the homogeneous nucleation theory is the 
assumption of formation of nucleating clusters within the 
initially homogeneous metastable state. The rate of conversion 
of the metastable state into a thermodynamically stable phase 
is called nucleation rate. According to Langer \cite{Lang67},
who has extended the classical nucleation theory 
(see, e.g. \cite{Fren46} and references herein) to field theories,
the rate of the relaxation of a metastable state is given by
\beq
\ds
I = \frac{\kappa}{2 \pi} \Omega_0\, \exp{(-\beta \Delta 
F_\ast)}\ ,
\label{f1}
\eeq
where $\kappa$ and $\Omega_0$ is a dynamical prefactor and a 
statistical prefactor, respectively, $\Delta F_\ast$ is the 
excess free energy of the cluster of critical size, $\beta 
\equiv (k_B T)^{-1}$, $k_B$ is the Boltzmann constant, and 
$T$ is the temperature of the system.

Let us consider a classical system with $N$ degrees of freedom
described by a set of $N$ collective coordinates 
$\xi_i,\ i = 1, \ldots , N$. The energy function $E(\{\xi\})$
of the system should have two locally stable states separated
by the energy barrier. The point with minimal energy on the
barrier $\{\xi^S\}$ is the so-called saddle point. 

The statistical prefactor $\Omega_0$ is a measure of the available
phase space volume of the saddle point region. If the fluctuation
corrections to the (mean field) excess free energy of the cluster
are absorbed into $\Delta F$, the statistical prefactor 
\cite{Lang67} can be written as
\beq
\ds
\Omega_0 = {\cal V} \left( \frac{2 \pi}{\beta |\lambda_1|}
\right)^{1/2}\ ,
\label{f2}
\eeq
where $\ds {\cal V} = V \left( \frac{8 \pi \sigma}{3 |\lambda_1|} 
\right)^{3/2}$ is the available volume of the saddle point subspace,
$V$ is the volume of the system, $\sigma$ is the 
surface tension, and $\lambda_1$ is the only negative eigenvalue of 
the matrix $\ds \frac{\partial^2 E}{\partial \xi_i \partial 
\xi_j}$ defined at the saddle point $\{\xi^S\}$. 

The dynamical prefactor $\kappa$ determines the growth rate of the 
critical cluster at the saddle point. 
For a relativistic system of particles, where the thermal
conductivity is absent because of the absence of the rest frame
defined by the baryon net charge, the dynamical prefactor has been
calculated \cite{CsKa92a} to be
\beq
\ds
\kappa = \frac{4 \sigma \left( \zeta + \frac{4}{3} \eta \right)}
{(\Delta \omega)^2\, R_\ast^3}\ .
\label{f3}
\eeq
Here $\zeta\ {\rm and}\ \eta$ is 
the shear and bulk viscosity, $\Delta \omega$ is the difference in
the enthalpy densities of the quark-gluon phase and hadronic phase,
and $R_\ast$ is the radius of the critical cluster.

We will explore a model QGP, produced in the collision of 
ultrarelativistic heavy ions, and subsequently rapidly quenched
into the metastable region below the critical temperature. Then the
first order phase transition should occur. It will be triggered by 
the creation of hadronic clusters as a result of the thermodynamic 
fluctuations. The Fisher droplet model \cite{Fish67} yields the 
change in the free energy of the system due to the formation of a 
spherical cluster of radius $R$
\beq
\ds
\Delta F(R) = - \frac{4\pi}{3}R^3\Delta p \, +\, 4\pi R^2 \sigma\,
 +\, 3 \tau_F \beta^{-1} \ln{\frac{R}{r_0}} \ .
\label{f4}
\eeq
Here $\Delta p = p_{h} - p_{qgp}$ is the difference of the pressures
inside and outside the cluster, $\tau_F = 2.2$ is the critical
exponent, and $r_0$ is the radius of the smallest cluster formed in
the system. The maximum of the $\Delta F(R)$ will be reached at 
the critical radius
\beq
\ds
R_\ast\, =\, \left( a_1\ +\ \sqrt{a_2^3 + a_1^2} \right)^{1/3}
  \ +\ \left( a_1\ -\ \sqrt{a_2^3 + a_1^2} \right)^{1/3}\ +\ a_3\ ,
\label{f4a}
\eeq
containing $\ds 
a_1 = \frac{9 \tau_F a_3}{4 \alpha^2}\,+\,a_3^3\ ,\ \ 
a_2 = -\,a_3^2\ ,\ \
a_3 = \frac{2 \sigma}{3 \Delta p}\ ,\ \
\alpha = \sqrt{4 \pi \sigma \beta}\ .$
Clusters whose radii are smaller than $R_\ast$ are shrinking,
clusters whose radii are larger than $R_\ast$ are growing,
and clusters of critical size are in metastable equilibrium.
It is convenient to parametrize the excess free energy of a cluster
by the similarity number 
$\ds \lambda_Z = R_\ast \sqrt{ 4 \pi \sigma \beta } $ \cite{BrZa95}
and reduced radius $r = R/\!R_\ast$ 
\beq
\ds
\beta \Delta F = - \left( \tau_F + \frac{2}{3} \lambda_Z^2 \right) 
r^3 + \lambda_Z^2 r^2 + 3\tau_F\,\ln \frac{r R_\ast}{r_0}\ .
\label{f5}
\eeq
Then for the only negative eigenvalue $\lambda_1$, associated with 
the instability of the critical cluster against growth or shrinking, 
we have
\beq
\lambda_1 = - ( 9 \tau_F + 2 \lambda_Z^2 ) \beta^{-1}\ .
\label{f7}
\eeq
Substitution of this result in Eq. (\ref{f2})
gives finally for the statistical prefactor
\beq
\ds
\Omega_0 = \frac{4 \sqrt{\pi}}{3 \sqrt{3}}\, \left( \frac{\lambda_Z}
{R_\ast} \right)^3 \, \frac{V}{(9 \tau_F + 2 \lambda_Z^2)^2}\ .
\label{f9}
\eeq
The result for the total nucleation rate $I$ reads
\beq
\ds
I = V \frac{8}{3 \sqrt{3 \pi}}\, 
\left( \frac{\lambda_Z}{R_\ast^2} \right)^3\,
\frac{\sigma \left( \zeta + \frac{4}{3} \eta \right)}
{(\Delta \omega)^2\, (9\tau_F + 2\lambda_Z^2)^2}\, 
\exp{\left( - \beta \Delta F_\ast \right)}\ ,
\label{f10}
\eeq
where $\ds \Delta F_\ast = - \frac{\tau_F}{\beta} \left( 1 - 3 \ln 
\frac{R_\ast}{r_0} \right) + \frac{\lambda_Z^2}{3 \beta}$ is the free 
energy of the critical cluster. The formula (\ref{f10}) is similar to,
but also different from the earliar expression \cite{CsKa92a}.
Near the critical temperature this rate drops since the critical
radius $R_\ast$ increases to infinity. Then the nucleation rate
rises quickly when the temperature falls, and the first order phase 
transition begins. 

From the definition of the nucleation rate, this is just a
number of viable clusters in volume $V$ passed through the
critical region per unit of time. If the total volume of the
hadronic phase is not large enough to maintain the temperature
of the expanding system at a constant level due to the release 
of the latent heat, the temperature will continue to decrease.

On the other hand, the critical radius drops very quickly with 
decreasing temperature. Therefore, the subcritical clusters
which started to dissolve (Fig. 1, upper panel) suddenly 
become of supercritical size with respect to the new critical 
radius $R_c^\prime$ of the system (Fig. 1, lower panel).
The general picture of the conversion of QGP into hadronic clusters
will be rather complex. Clusters large enough will grow, the 
smallest ones will continue to shrink, while the clusters of medium 
size will grow or shrink depending on temperature variations and 
current critical radii in the system.

It is worth noting that the model presented above is valid for the
hadronization of thermalized QGP produced in a large volume (of about
500$\,$fm$^3$ or more). 
For smaller systems, the finite size effects \cite{SSG97} lead to the 
shift of the critical temperature and rounding of the phase transition, 
but the discussion of these problems lies out of scope of the present 
paper.

At given temperature $T$, the size distribution of clusters in the
droplet model with the curvature energy is found to be \cite{BrZa96}
\beqar
\ds
f(R) &=& \frac{I}{\sqrt{2 \pi (9 \tau_F + 2 \lambda_Z^2)}} 
\exp \left[ \frac{9 \tau_F + 2 \lambda_Z^2}{2}\, (r - 1)^2 \right] \\
&\times & 
\int_r^\infty a \, \left[ 3 \tau_F (a^2 + a + 1) + 2 \lambda_Z^2 a^2
\right]\, \exp \left[ - \frac{9 \tau_F + 2 \lambda_Z^2}{2}\,
(a - 1)^2 \right] da \ .
\nonumber
\label{f11}
\eeqar
Hadronic matter in our calculations is represented by a discrete 
spectrum of spherical bubbles starting from $r_0 = 1\,$fm.

Within the framework of the bag model the equations of state for 
two-flavour QGP and for hadronic gas are
\beqar
\ds
p_{qgp} &=& \frac{37}{90} \pi^2 T^4 - B \ , \\
\label{f12}
p_{h} &=& \frac{1}{30}  \pi^2 T^4 \ .
\label{f13}
\eeqar
With the bag constant $B^{1/4} = 235\,$MeV this gives us the 
critical temperature $T_c = 169$ MeV. 

The Bjorken model of longitudinal expansion \cite{Bjor83} yields for 
the evolution of the energy density $e$
\beq
\ds
\frac{d e}{d \tau} = - \frac{e + p}{\tau} \ .
\label{f14}
\eeq
The energy density of the whole system, as well as the enthalpy,
is a linear combination of the energy density of the plasma $e_{qgp}$
in the part of volume occupied by plasma, $\nu_{qgp} = 
V_{qgp}/V_{tot}$, and the energy density of hadronic phase $e_{h}$ in 
the rest of the volume, $\nu_{h} = 1 - \nu_{qgp} $, 
\beq
\ds
e = e_{qgp} \nu_{qgp} + (1 - \nu_{qgp}) e_{h} \ .
\label{f15a}
\eeq

To determine the viscous term in Eq. (\ref{f3}), note that the bulk
viscosity is much smaller as compared with the shear viscosity, which 
has been derived in leading logarithmic order of QCD \cite{Heis94} 
for a QGP with two flavours as
\beq
\ds
\eta = \frac{1.29\, T^3}{\alpha_S^2\, \ln{\left( 1/\alpha_S 
\right)}}\ .
\label{f16}
\eeq
Here $\alpha_S = 0.23$ is the strong coupling constant.

The last important parameter for our analysis is the value of the
surface tension $\sigma$. Recent lattice QCD calculations 
\cite{Iwa94,Kar96} predict 
$0.01 < \sigma/T_c^3 < 0.1$, i.e. $1.25 < \sigma < 12.5$ 
MeV/fm$^2$ for the given $T_c$.  We will use $\sigma = 2$ and 5 
MeV/fm$^2$ in our simulations. 

Following \cite{CsKa92b,CKKZ93}, our scenario of longitudinal 
expansion presumes 
that the system reaches the critical temperature at the time 
$\tau_{cr} = 8\, \tau_{init} = 3\,$fm/$c$. From Fig. 2 it is 
apparent that the large amount of latent heat, released during 
the conversion process, is sufficient to prevent the expanding
system from the supercooling below $3-6\,\%$. 

To investigate the
effect of the dissipative processes on the course of the phase
transition, Fig. 2 shows also results of calculations, in which 
Eq. (\ref{f14}) has been replaced \cite{DaGy85} by
\beq
\ds
\frac{d e}{d \tau} = - \frac{e + p}{\tau} + \frac{\zeta + 
\frac{4}{3}\eta} {\tau^2}\ .
\label{f17}
\eeq
One can see that the supercooling of a viscous plasma is weaker 
than that of an ideal QGP. This circumstance delays the homogeneous
nucleation by only about $2-3$ fm/$c$. Creation of hadronic bubbles 
is the main mechanism of the QGP conversion at the earlier stage of 
the transition. Apart from the first fm/$c$ of cooling the mixed 
system is reheated continuously, and the temperature
approaches the critical one.

This occurs up to $\approx 6$ fm/$c$, then the behaviour of the 
system changes drastically: diffusion growth of the "old" bubbles
(Fig. 3, upper panels) dominates the creation of new bubbles, which
is practically turned off.
When $T$ is close to $T_c$, the growth process stops, but it starts 
again immediately after a small extra-cooling of the system. 
This is a highly unstable region. Even an insignificant rise of
the temperature is sufficient for the system to hit the critical 
point, then the homogeneous nucleation scheme is no longer 
relevant. Therefore, the spinodal decomposition or 
the percolation scenario may be appropriate to describe the 
strongly non-equilibrium hadronization of the rest of the QGP, 
which constitutes about $30-35\%$ of the total volume.

The size distribution of the hadronic bubbles at different stages of
the phase transition is shown in lower panels of Fig. 3. We see that 
the initially
broad plateau of the distribution function becomes narrower with the
rise of the radius due to the increase of the number of bubbles per 
unit of radial interval. The distribution functions reach their 
maximum values with a radius of about $3 \leq r \leq 4.5\,$fm. 
At the very beginning of the nucleation almost 
all bubbles enlarge their volumes. Then the bubbles of 
$r \leq 1.5\,$fm dissolve because their radii are smaller 
than the critical one.
In contrast, the bubbles of $r \geq 4\,$fm will stop to grow only
if the temperature of the system will be very close to the critical
temperature. Bubbles of size $1.5 \leq r \leq 4\,$fm are either
growing or shrinking due to the variations of the critical radius in 
the system. This leads to the appearance of the irregularities in the
spectrum of small and intermediate-size bubbles (Fig. 3, right lower
panel). 

As a consequence, there will be no unique source of pion emission at 
the freeze-out. The individual pairs of pions are coming from
sources of $r \approx 1.5-2\,$fm, while the major part of pions should
come from relatively large sources of $3 \leq r \leq 4.5\,$fm.
This can be checked by the analysis of the particle correlation data
and by the rapidity distributions, if the hadronic clusters will be
well separated along the expansion axis; this problem has to be
investigated.

In conclusion, we study two-flavour QGP undergoing a first order phase 
transition during the longitudinal expansion. The nucleation rate is 
derived via the new scaling parameter $\lambda_Z$. 
It is shown that the distribution of hadronic clusters
in size plays an essential role in the dynamics of plasma
hadronization near the critical temperature: there is a significant
variation of the value of critical cluster in the system. The
supercooling of QGP is found to be relatively moderate, $3-6\%$
only, therefore the application of the homogeneous nucleation
scenario seems to be quite reasonable.

The completion time of the transition varies from 6 to 10 fm/$c$.
It is strongly dependent on the absolute value of the interfacial 
energy. The weaker the surface tension $\sigma$ of the interface is,
the faster the hadronization will be, and vice versa. 
Since recent lattice QCD results \cite{Kar96} favour lower values of 
$\sigma$, it is likely that the QGP hadronizes
within first eight fm/$c\/$'s, i.e. very close to the currant 
estimations of the freeze-out time scale based on HBT data and
microscopic models. Thus the completion time of the homogeneous
nucleation stage is about $5-10$ times shorter than those of the
standard nucleation scheme obtained before \cite{CsKa92b,CKKZ93}.

The size distribution of the hadronic clusters indicates that most of 
the pions are emitted from sources of about $3 \leq r \leq 4.5\,$fm. 
This signal may be investigated experimentally by means of the 
particle interferometry and, probably, by the rapidity spectra of 
secondaries on an event-by-event basis.

{\bf Acknowledgements.} 
We thank M.I. Gorenstein for useful discussions.
L.B. and E.Z. are grateful to the Institute for Theoretical Physics, 
University of Frankfurt for the warm and kind hospitality. 
L.B. acknowledges support of the Alexander-von-Humboldt Foundation.

\newpage

\begin{figure}[htp]
\centerline{\epsfysize=18cm \epsfbox{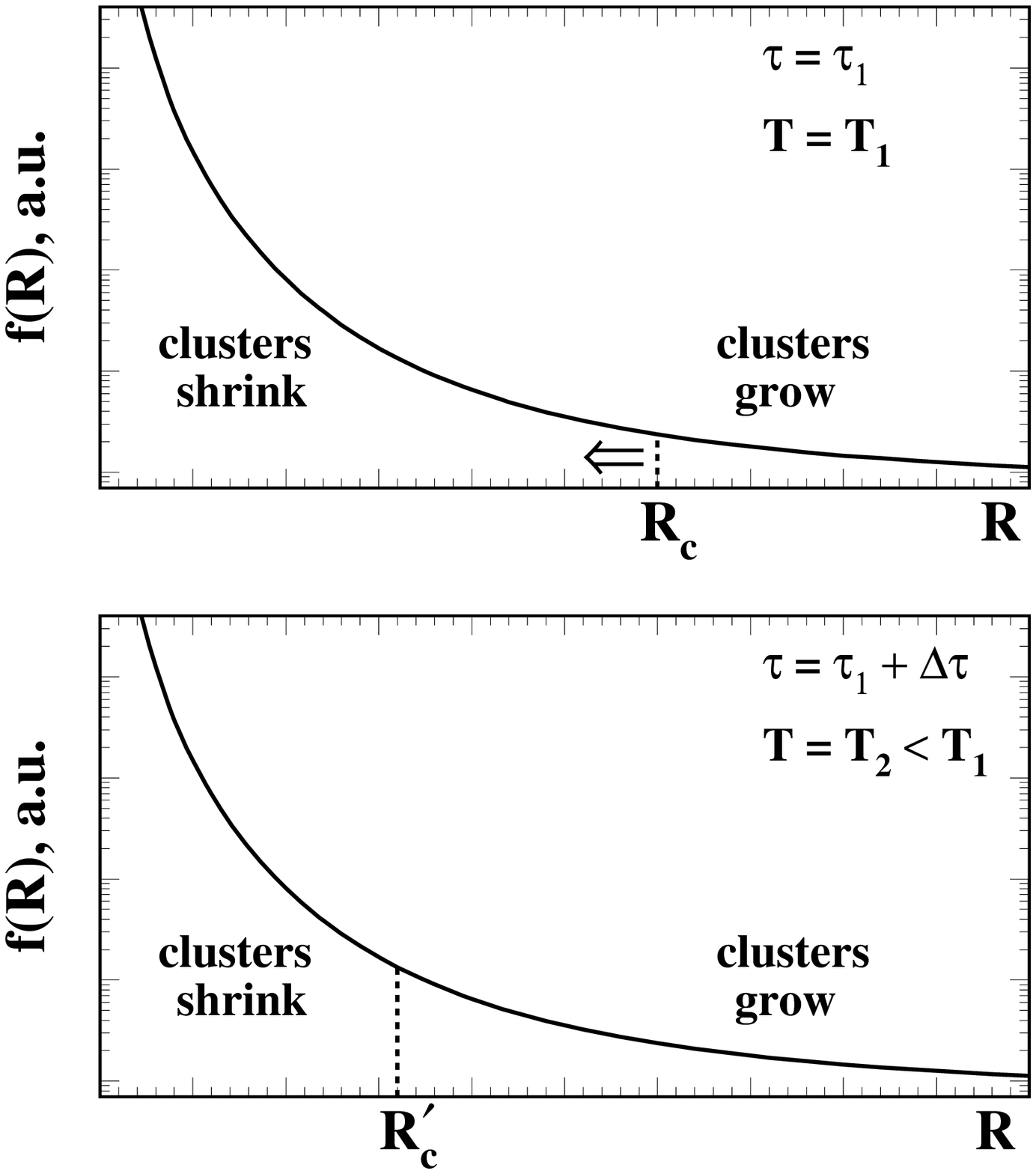}}
\caption{
Distribution of clusters as a function of their radius $R$.
Due to the rapid falloff of the critical radius with temperature,
the clusters of subcritical sizes (upper panel) formed at the moment
$\tau_1$ become supercritically sized (lower panel) at 
$\tau_1 + \Delta \tau$.  Consequently, the amount of matter converted
into hadrons increases.}
\label{fig1}
\end{figure}

\begin{figure}[htp]
\centerline{\epsfysize=18cm \epsfbox{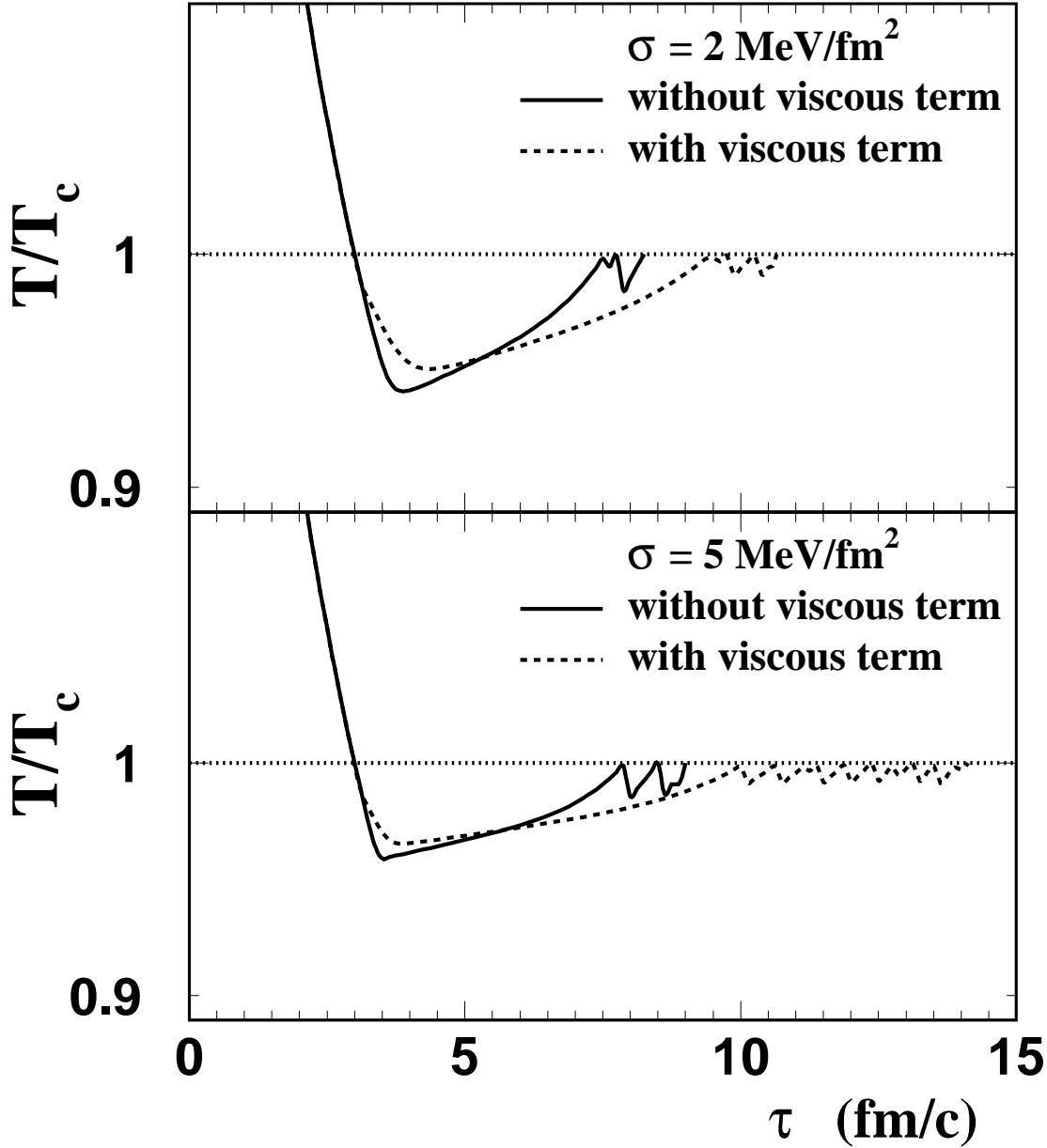}}
\caption{
The temperature as a function of the proper time for 
the phase transition QGP $\rightarrow$ hadrons during the 
longitudinal expansion. Upper and lower panel represents the results 
of calculations with $\sigma = 2$ and 5 MeV/fm$^2$, respectively.
The solid/dashed curves correspond to calculations 
with Eq.~(13)/Eq.~(16).}
\label{fig2}
\end{figure}

\begin{figure}[htp]
\centerline{\epsfysize=18cm \epsfbox{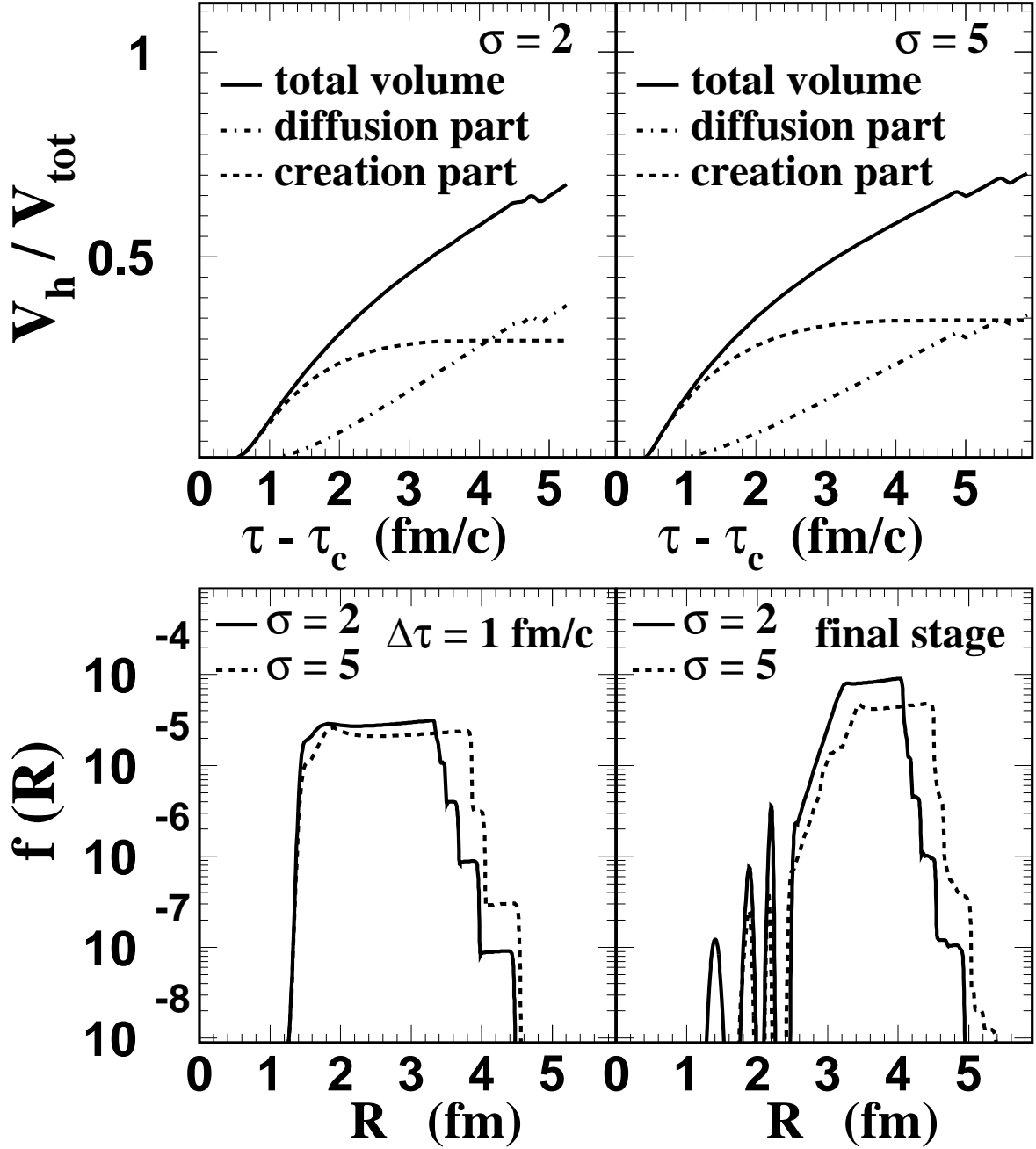}}
\caption{
{\bf Upper :} The part of the QGP volume converted into 
hadrons for system with $\sigma = 2$ (left panel) and 5 MeV/fm$^2$ 
(right panel). 
Solid lines correspond to the total volume fractions of hadrons,
dashed lines denote the increase of the hadronic volume due to the
creation of new bubbles, and dash-dotted lines indicate the
enlargement of the hadronic bubbles due to diffusion.
{\bf Lower :}
Size distribution of the hadronic bubbles at $\Delta \tau =
1\,$fm/$c$ after the beginning of the nucleation (left panel), 
and at the freeze-out (right panel) with $\sigma = 2$ (solid curves) 
and 5 MeV/fm$^2$ (dashed curves).}
\label{fig3}
\end{figure}

\end{document}